\documentclass[aps,a4paper,prd,groupedaddress,preprintnumbers,nofootinbib,preprint,floatfix]{revtex4}

\makeatother

\usepackage[english]{babel}
\usepackage{amsmath}
\usepackage{amssymb}
\usepackage{bbm}
\usepackage{hyperref}
\usepackage{physics}
\usepackage{accents}
\usepackage[utf8]{inputenc}
\usepackage{setspace}
\usepackage{graphicx,caption}
\usepackage[normalem]{ulem}
\usepackage{xcolor}

\usepackage{hyperref}
\hypersetup{breaklinks=true, colorlinks=false, linkcolor=black, urlcolor=black}
\usepackage{breakurl}

\newcommand{\be}{\begin{equation}}
\newcommand{\ee}{\end{equation}}
\newcommand{\ba}{\begin{eqnarray}}
\newcommand{\ea}{\end{eqnarray}}
\newcommand{\bs}{\begin{subequations}}
\newcommand{\es}{\end{subequations}}

\newtheorem{theorem}{Theorem}[section]
\newtheorem{corollary}{Corollary}[theorem]

\newcommand{\AddrIST}{Universidade de Lisboa,
  Instituto Superior T\'ecnico, CFTP, 1049-001~Lisboa, Portugal}

\begin{document}


\title{Theorem on vacuum stability}

\author{Andr\'e Milagre}%
\email[E-mail: ]{andre.milagre@tecnico.ulisboa.pt}
\affiliation{\AddrIST}
\author{Lu\'\i s Lavoura}
\email[E-mail: ]{balio@cftp.tecnico.ulisboa.pt}
\affiliation{\AddrIST}
\today

\begin{abstract}
  We consider an extension of the Standard Model
  with one or more scalar multiplets beyond the Higgs doublet $\Phi$.
  The additional scalar multiplets are supposed to carry
  arbitrary hypercharges
  (the precise conditions
  on the scalar potential are less stringent).
  We prove that,
  in such a model,
  if the field configuration
  where only $\Phi$ has a nonzero vacuum expectation value (VEV)
  is a local minimum of the potential,
  then it has a lower value of the potential than any extremum
  where both $\Phi$ and other scalar multiplets have nonzero VEVs.
\end{abstract}

\maketitle

\section{Introduction}

In 2019 Ferreira and Gon\c calves (FG)~\cite{goncalves}
considered the extension of the Standard Model
where the scalar sector consists of one $SU(2)$ doublet $\Phi$
and one $SU(2)$ triplet $\Delta$:
\be
\Phi = \left( \begin{array}{c} \phi_1 \\ \phi_2 \end{array} \right),
\qquad
\Delta  = \left( \begin{array}{cc}
  - \delta_2 \left/ \sqrt{2} \right. & \delta_1 \\
  - \delta_3 & \delta_2 \left/ \sqrt{2} \right.
\end{array} \right),
\ee
where $\phi_1$,
$\phi_2$,
$\delta_1$,
$\delta_2$,
and $\delta_3$ are complex scalar fields.
They assumed that the scalar potential is
\bs
\label{1}
\ba
V &=& m^2\, \Phi^\dagger \Phi
+ M^2\, \mathrm{tr} \left( \Delta^\dagger \Delta \right)
+ \lambda_1 \left( \Phi^\dagger \Phi \right)^2
+ \lambda_2 \left[ \mathrm{tr} \left( \Delta^\dagger \Delta \right) \right]^2
\\ & &
+ \lambda_3\, \mathrm{tr} \left( \Delta^\dagger \Delta \Delta^\dagger \Delta
\right)
+ \lambda_4 \left( \Phi^\dagger \Phi \right)
\mathrm{tr} \left( \Delta^\dagger \Delta \right)
+ \lambda_5\, \Phi^\dagger \Delta \Delta^\dagger \Phi,
\ea
\es
which is invariant under $\Phi \to U \Phi,\
\Delta \to U \Delta\, U^\dagger$,
where $U$ is an $SU(2)$ matrix.
A crucial feature of the potential in Eq.~\eqref{1} is that it is also invariant
under \emph{independent} rephasings of $\Phi$ and $\Delta$,
\textit{viz.}\ both under $\Phi \to \Phi\, \exp \left( i \varkappa \right)$
and under $\Delta \to \Delta\, \exp \left( i \varpi \right)$.
FG also assumed that there is a stationary point of $V$
where either one or both fields in $\Phi$
have nonzero vacuum expectation values (VEVs)
while all the fields in $\Delta$ have zero VEVs---we call this a
`type-I extremum'---and that this stationary point is a \emph{local minimum},
\textit{i.e.}\ that it leads to all the physical particles---\textit{i.e.}\
all but the Goldstone bosons---having \emph{positive} masses-squared.
Under these assumptions,
FG proved that that type-I extremum has a lower value of $V$
than \emph{any} stationary point of $V$
where \emph{both} $\Phi$ and $\Delta$ have nonzero VEVs---we call this a
`type-III extremum'.
FG emphasized,
though,
that an extremum where only fields in $\Delta$ have nonzero VEVs
while both $\phi_1$ and $\phi_2$ have zero VEVs---we call this
a `type-II extremum'---may have
a lower value of $V$ than the type-I extremum.

One may exploit this observation and require the parameters 
of a scalar potential to be such that the global minimum 
corresponds to a vacuum configuration in which only
the neutral component of the Higgs doublet acquires VEV.
Such theoretical constraints have been
used to build consistent minimal extensions of the Standard Model.
For instance, the authors of Refs.~\cite{Ferreira:2004yd,Barroso:2005sm} 
have analyzed the vacua of the most general two-Higgs-doublet model (2HDM)
and proved that if a charge-preserving minimum exists,
then any extremum that spontaneously breaks 
either charge-conservation, $C$, or $CP$ symmetries
is necessarily a saddle point lying above that minimum.
Stability conditions of neutral vacua have also been derived for 
the real \cite{Belanger:2012zr} and complex \cite{Ferreira:2016tcu} 
scalar singlet models,
the left-right symmetric model \cite{BhupalDev:2018xya}, 
the Georgi--Machacek model \cite{Hartling:2014zca,Azevedo:2020mjg},
the Next-to-Minimal 2HDM (N2HDM) \cite{Ferreira:2019iqb},
some symmetry-constrained realizations of the
Three-Higgs-Doublet Model (3HDM)~\cite{Ivanov:2014doa,Faro:2019vcd,Hernandez-Sanchez:2020aop,Boto:2024tzp},
and the $N$-Higgs-Doublet Model
(NHDM)~\cite{Barroso:2006pa,Nishi:2007nh,Ivanov:2010ww}.
Notably,
Ref.~\cite{Barroso:2006pa} has asserted that,
in the NHDM,
the depth of the potential at a stationary point that breaks a given symmetry, 
relative to a neutral minimum, 
is related to the mass-squared matrix of the scalar particles
directly linked with that symmetry.
We recover that result
in the present paper.

In particular, in Dark Matter models, the dark matter scalar 
multiplets must be VEV-less
in order not to break their stabilizing symmetries. This requires
vacuum stability conditions, \textit{viz}. one should make sure that the
desired vacuum structure corresponds to the true global minimum
of the potential.
The present paper gives one step that may be useful for this
purpose.

In section~\ref{sec:theorem} of this paper we state a 
theorem that generalizes FG's result
to the case where $\Delta$ is an $SU(2)$ multiplet---or,
indeed,
a set of $SU(2)$ multiplets---with \emph{any} isospin.
In section \ref{sec:proof} we prove the theorem.
We summarize our findings in section~\ref{sec:conclusions}.
We use appendix \ref{sec:appA} to apply
our theorem to the
$U(1)$-symmetric 2HDM~\cite{Branco:2011iw}.

\section{\label{sec:theorem}The theorem} 

We consider an extension of the SM with one $SU(2)$
scalar doublet $\Phi$
and $N$ $SU(2)$ scalar multiplets $\Psi_k$ ($k= 1, \ldots, N$).
We write the field components of $\Phi$
and of its charge-conjugate $SU(2)$ doublet $\widetilde{\Phi}$ as
\be
\Phi = \left( \begin{array}{c} \phi_1 \\ \phi_2 \end{array} \right),
\quad
\widetilde{\Phi} = \left( \begin{array}{c} \phi_2^* \\ - \phi_1^*
\end{array} \right).
\ee
Each multiplet $\Psi_k$ has isospin $J_k$ and dimension $n_k = 2 J_k + 1$.
Note that the isospins and hypercharges of the multiplets $\Psi_k$
are constrained by partial wave unitarity bounds, 
as found in Ref.~\cite{logan,Milagre:2024wcg}.
We label the field components of $\Psi_k$ as $\psi_{k, I}$,
where $I$  is the third component of isospin,
ranging from $-J_k$ to $J_k$ in discrete steps $\Delta I = 1$.
Explicitly,
\be
\Psi_k = \left( \begin{array}{c} 
\psi_{k, J_k}\\ 
\psi_{k, J_k-1} \\ 
\psi_{k, J_k-2} \\
\vdots \\
\psi_{k, 1-J_k} \\
\psi_{k, -J_k} 
\end{array} \right),
\quad
\widetilde \Psi_k = \left( \begin{array}{c} 
\psi_{k, -J_k}^*\\ 
-\psi_{k, 1-J_k}^* \\ 
\psi_{k, 2-J_k}^* \\
\vdots \\
\left( -1 \right)^{n_k} \psi_{k, J_k-1} ^*\\
- \left( -1 \right)^{n_k} \psi_{k, J_k}^* 
\end{array} \right).
\ee
The scalar fields $\phi_1$,
$\phi_2$,
and $\psi_{k, I}$ are complex and their third components of isospin are
\be
I \left( \phi_1 \right) = \frac{1}{2}, \quad
I \left( \phi_2 \right) = - \frac{1}{2}, \quad
I \left( \psi_{k, I} \right) = I,
\ee
respectively.

Let $V$ be the scalar potential of $\Phi$ and the $\Psi_k$.
We make the following assumptions:
\begin{enumerate}
\item $V$ is invariant under the gauge symmetry $SU(2) \times U(1)$.
\item $V$ is renormalizable,
  \textit{i.e.}\ it does not contain any terms with more than four fields.
\item $V$ does not contain any linear terms.
  This is a trivial consequence of $SU(2) \times U(1)$-invariance
  in the case of a $\Psi_k$ that is not $SU(2) \times U(1)$-invariant.
  Else,
  it may be achieved,
  for instance,
  through an additional symmetry $\Psi_k \to - \Psi_k$
  for each $\Psi_k$ which is $SU(2) \times U(1)$-invariant;
  or else by collecting all such $\Psi_k$ in a non-trivial representation
  of some non-Abelian horizontal-symmetry group.
\item $V$ does not contain any trilinear terms.
  This may be a consequence either of the $SU(2) \times U(1)$ symmetry
  or of some additional (non-gauge) symmetry.
  For instance,
  if a $\Psi_k$ has half-integer $J_k$,
  then $SU(2)$-invariance forbids any trilinear term
  only with $\Phi$ and that $\Psi_k$.
  Another instance:
  if a $\Psi_k$ is a triplet of $SU(2)$,
  then $SU(2) \times U(1)$-invariance forbids any trilinear term
  of $\Phi$ and that $\Psi_k$
  if $\Psi_k$ has hypercharge different from either $0$ or $\pm 1$;
  or else,
  there may be an additional symmetry $\Psi_k \to - \Psi_k$
  that forbids the trilinear coupling $\Phi \Phi \Psi_k$.
\item $V$ does not contain any bilinear terms with two different multiplets.
  These may be terms of the form either $\Phi \otimes \Psi_k$ 
  or $\Psi_k \otimes \Psi_{k^\prime}$ with $k \neq k^\prime$;
  they may be eliminated either by the gauge symmetry
  (for instance,
  if $\Phi$ and all the $\Psi_k$ have different isospins),
  or by some additional symmetry, like a separate rephasing symmetry
  $\Psi_k \to \Psi_k \exp \left( i \vartheta_k \right)$ for each $k$.
\item $V$ also does not contain any bilinear terms
  of the form $\Psi_k \otimes \Psi_k$,
  but for
  \be
  F_k \equiv \sum_{I=-J_k}^{J_k} \left| \psi_{k, I} \right|^2.
  \label{psi_1}
  \ee
  For instance,
  in the case of the triplet $\Delta$ of the previous section,
  there might be a term $\mathrm{tr} \left( \Delta^2 \right) =
  \left( \delta_2 \right)^2 - 2 \delta_1 \delta_3$ in the scalar potential;
  we assume that some symmetry eliminates that term.
  Once again,
  a $U(1)$ symmetry
  $\Psi_k \to \Psi_k \exp \left( i \vartheta_k \right)$ for each $k$
  is the handiest way to eliminate such terms.
\item The only $SU(2) \times U(1)$-invariant terms in $V$
  with both $\Phi$ and $\Psi_k$ are constructed out of
  \be
  \left(
  {\Phi \otimes \widetilde{\Phi} \otimes \Psi_k \otimes \widetilde{\Psi}_k}
  \right).
  \label{pppp}
  \ee
  From the product of $\Phi$ and $\widetilde{\Phi}$
  we get both a singlet and a triplet of $SU(2)$:
  \bs
  \ba
  F_0 \equiv \left( \Phi \otimes \widetilde{\Phi} \right)_\mathbf{1} &=&
  \left| \phi_1 \right|^2 + \left| \phi_2 \right|^2,
  \label{phi_1} \\
  \left( \Phi \otimes \widetilde{\Phi} \right)_\mathbf{3} &=&
  \left( \begin{array}{c}
    \displaystyle{\phi_1 \phi_2^*} \\*[2mm]
    \displaystyle{\frac{\left| \phi_2 \right|^2
        - \left| \phi_1 \right|^2}{\sqrt{2}}}
    \\*[2mm]
    \displaystyle{- \phi_1^* \phi_2}
  \end{array} \right)\label{phi_3}.
  \ea
  \es
  As a consequence,
  only two $SU(2)$ invariants may be built from the expression~\eqref{pppp}:
  \be
  \left( \Phi \otimes \widetilde{\Phi}
  \otimes \Psi_k \otimes \widetilde{\Psi}_k \right)_\mathbf{1}
  = \left( \Phi \otimes \widetilde{\Phi} \right)_\mathbf{1}
  \left( \Psi_k \otimes \widetilde{\Psi}_k \right)_\mathbf{1}
  \, \oplus \,
  \left[
    \left( \Phi \otimes \widetilde{\Phi} \right)_\mathbf{3}
    \otimes
    \left( \Psi_k \otimes \widetilde{\Psi}_k \right)_\mathbf{3}
    \right]_\mathbf{1},
  \ee
  where $\left( \Psi_k \otimes \widetilde{\Psi}_k \right)_\mathbf{1}$
  is the quantity $F_k$ given in Eq.~\eqref{psi_1}.
  Using Clebsch–Gordan coefficients in the usual manner,
  we derive
  \be
  F_{0k} \equiv \left[\left( \Phi \otimes \widetilde{\Phi} \right)_\mathbf{3}
    \otimes
    \left( \Psi_k \otimes \widetilde{\Psi}_k \right)_\mathbf{3} \right]_\mathbf{1}
  = \frac{\left| \phi_1 \right|^2 - \left| \phi_2 \right|^2}{2}\,
  \sum_{I=-J_k}^{J_k} I \left| \psi_{k, I} \right|^2
  + \frac{z+z^*}{2},
  \label{13b}
  \ee
  where
  \be
  z = \phi_1 \phi_2^* \sum_{I=1-J_k}^{J_k}
  \psi_{k, I}^*\, \psi_{k, I-1}\ \sqrt{J_k^2 - I^2 + J_k + I}.
  \ee
\end{enumerate}
Putting all these conditions together,
one concludes that
\bs
\label{SP}	
\ba
V &=&
\mu_0^2\, F_0 + \frac{\lambda_{00}}{2}\, F_0^2 
+ \sum_{k=1}^N   \left( \lambda_{0k} F_0 F_k
+ \lambda_{0k}^\prime F_{0k} \right)
\label{line15a}\\
& & +\  
\sum_{k=1}^N   \mu_{k}^2 F_k 
+\frac{1}{2}\, \sum_{k=1}^N\, \sum_{k^\prime=1}^N
\lambda_{kk^\prime} F_{i} F_{k^\prime} 
\  + \  \textrm{other terms quartic in the $\Psi_k$},
\ea
\es
where $F_0$,
$F_k$,
and $F_{0k}$ have been defined in Eqs.~\eqref{phi_1},
\eqref{psi_1},
and~\eqref{13b},
respectively;
$\mu_0^2$ and $\mu_k^2$ are real coefficients
with squared-mass dimension;
and $\lambda_{00}$,
$\lambda_{0k}$,
$\lambda_{0k}^\prime$,
and $\lambda_{k k^\prime}$ are real dimensionless coefficients
(note that $\lambda_{k k^\prime} = \lambda_{k^\prime k}$
for all pairs $k \neq k^\prime$).\footnote{In the case where $\Psi_1$
is an $SU(2)$ triplet and there are no $\Psi_k$ with $k > 1$,
one may compare the potential~\eqref{SP} with FG's one in Eq.~\eqref{1}.
One sees that $\mu_0^2 = m^2$,
$\mu_1^2 = M^2$,
$\lambda_{00} = 2 \lambda_1$,
$\lambda_{11} = 2 \lambda_2$,
$\lambda_{01} = \lambda_4 + \lambda_5 / 2$,
and $\lambda_{01}^\prime = \lambda_5$.
Moreover,
the ``other terms quartic in the $\Psi_k$'' of Eq.~\eqref{SP}
consist just of the term with coefficient $\lambda_3$ in Eq.~\eqref{1}.}

The scalar potential may have various extrema.
We classify those extrema in the following way:
\begin{itemize}
\item The type-0 extremum has
  \be
  \label{3}
  \left\langle \Phi \right\rangle = 0,
  \quad
  \forall k\, \left\langle \Psi_k \right\rangle = 0.
  \ee
\item A type-I extremum has
  \be
  \label{4}
  \left\langle \Phi \right\rangle \neq 0,
  \quad
  \forall k\, \left\langle \Psi_k \right\rangle = 0.
  \ee
\item A type-II extremum has
  \be
  \label{5}
  \left\langle \Phi \right\rangle = 0,
  \quad
  \exists\, \left\langle \Psi_k \right\rangle \neq 0.
  \ee
\item A type-III extremum has
  \be
  \label{6}
  \left\langle \Phi \right\rangle \neq 0,
  \quad
  \exists\, \left\langle \Psi_k \right\rangle \neq 0.
  \ee
\end{itemize}
The four types of extrema are qualitatively different
and one cannot continuously transition from one to another.
In Eqs.~\eqref{3}--\eqref{6},
$\langle \rangle$ denotes a field's vacuum expectation value (VEV).

We may now state
\begin{theorem}
  A local type-I minimum of $V$ in Eq.~\eqref{SP}
  has a lower expectation value of the
  potential than any type-0 or type-III extremum.
\end{theorem}
%
\begin{corollary}
  Either the global minimum of the potential is that type-I local minimum,
  or it is a type-II local minimum (if there is any).
\end{corollary}

\section{\label{sec:proof}Proof}

\subsection{\label{subsec:SP} The scalar potential at an extremum}

Given the structure of $V$,
one may rewrite Eq.~\eqref{SP} as
\be
V = L^T X + \frac{1}{2}\, X^T M X 
+ \sum_{k=1}^N \lambda_{0k}^\prime F_{0k}
+ \ \textrm{other terms quartic in the $\Psi_k$},
\label{17}
\ee
where
\bs
\ba
X &=& \left( \begin{array}{cccc} F_0, & F_1, & \ldots, & F_N
\end{array} \right)^T,
\label{X} \\
L &=& \left( \begin{array}{cccc} \mu^2_0, & \mu^2_1, & \ldots, & \mu^2_N
\end{array} \right)^T,
\label{L} \\
M &=&
\left( 
\begin{array}{cccc} 
\lambda_{00} & \lambda_{01} & \cdots & \lambda_{0N} \\
\lambda_{01} & \lambda_{11}, &	\cdots	&	\lambda_{1N} \\
\vdots	& \vdots	&	\ddots	&	\vdots \\
\lambda_{0N}  &	\lambda_{1N}	 &	\cdots	& \lambda_{NN}
\end{array} 
\right).
\label{M}
\ea
\es
Note that the matrix $M$ is symmetric.

In Eq.~\eqref{17} one sees that
\bs
\ba
\frac{\partial V}{\partial X}
&=& \left( \begin{array}{cccc}
  \displaystyle{\frac{\partial V}{\partial F_0},} &
  \displaystyle{\frac{\partial V}{\partial F_1},} &
  \ldots, &
  \displaystyle{\frac{\partial V}{\partial F_N}}
\end{array} \right)^T
\\ &=& L + X^T M,
\label{dVdX} \\
\frac{\partial V}{\partial F_{0k}} &=& \lambda_{0k}^\prime.
\label{dVdF0i}
\ea
\es

The potential is the sum of two homogeneous functions\footnote{
A function $f : \mathbb{C}^n \to \mathbb{R}$ 
is said to be \textit{homogeneous of degree} 
$d \in \mathbb{Z}$ if
\be 
f(\alpha x_1, \ldots, \alpha x_n) = \alpha^d f(x_1, \ldots, x_n),
\ee
for all $\alpha \in \mathbb{R}$ and all 
$(x_1, \ldots, x_n) \in \mathbb{R}^n$~\cite{homog_euler}.}
of the scalar fields: a homogeneous function of degree 2
(which we call $V_{(2)}$),
and a homogeneous function of degree 4
(which we call $V_{(4)}$).
Indeed,
\bs
\ba
V &=& V_{(2)} + V_{(4)},
\\
V_{(2)} &=& L^T X,
\\
V_{(4)} &=&  \frac{1}{2}\, X^T M X 
+ \sum_{k=1}^N \lambda_{0k}^\prime F_{0k}
+ \textrm{other terms quartic in the $\Psi_k$}.
\ea
\es

We define a vector $\varphi$ containing all the fields in the theory:
\begin{equation}
  \varphi=\left( \begin{array}{cccccccc} 
  \phi_1, 
  & \phi_2,
  & \psi_{1, J_1},
  & \ldots,
  & \psi_{1, -J_1},
  & \psi_{2, J_2},
  & \ldots,	
  & \psi_{N, -J_N}
    \end{array} \right)^T.
\end{equation}
It follows from Euler's Theorem for homogeneous functions that~\cite{homog_euler}
\bs
\ba
\sum_\alpha \left( \varphi_\alpha\,
\frac{\partial V_{(2)}}{\partial \varphi_\alpha}
+ \varphi_\alpha^*\,
\frac{\partial V_{(2)}}{\partial \varphi_\alpha^*} \right) &=& 2\, V_{(2)},\\
\sum_\alpha \left( \varphi_\alpha\,
\frac{\partial V_{(4)}}{\partial \varphi_\alpha}
+ \varphi_\alpha^*\,
\frac{\partial V_{(4)}}{\partial \varphi_\alpha^*} \right) &=& 4\, V_{(4)},
\ea
\es
hence
\begin{equation}
\sum_\alpha \left( \varphi_\alpha\,
\frac{\partial V}{\partial \varphi_\alpha}
+ \varphi_\alpha^*\,
\frac{\partial V}{\partial \varphi_\alpha^*} \right)
= 2\, V_{(2)} + 4\, V_{(4)}.
\label{homogen}
\end{equation}

At any extremum of $V$,
the fields have a certain vacuum configuration $\varphi = \overline \varphi$.
Let $\overline V$,
$\overline V_{(2)}$,
and $\overline V_{(4)}$
denote the values of $V$,
$V_{(2)}$,
and $V_{(4)}$,
respectively,
when $\varphi = \overline \varphi$.
The stationarity conditions imply that, for every field $\varphi_\alpha$,
\be
\left.\frac{\partial V}{\partial \varphi_\alpha}
\right|_{\varphi = \overline \varphi}
=
\left.\frac{\partial V}{\partial \varphi_\alpha^*}
\right|_{\varphi = \overline \varphi}
= 0.
\label{stat_cond}
\ee
Therefore,
at any extremum,
Eq.~\eqref{homogen} reads
\be
2\, \overline{V}_{(2)} + 4\, \overline{V}_{(4)} = 0,
\ee
which implies $\overline{V}_{(4)} = - \left. \overline{V}_{(2)} \right/ 2$
and therefore
\be
\label{mc09}
\overline V = \overline V_{(2)} + \overline V_{(4)}
= \frac{\overline{V}_{(2)}}{2}
= \frac{L^T \, \overline X}{2},
\ee
where
\be
\overline X \equiv \left. X \right|_{\varphi = \overline \varphi}.
\ee

\subsection{The type-0 extremum}

At the type-0 extremum no field has a VEV,
hence $\overline{X} \equiv \overline{X}_0$ is a null vector.
If $\overline{V}_0$ is the value of the potential
at the type-0 extremum,
it follows from Eq.~\eqref{mc09} that
\be
\overline{V}_0 = \frac{L^T \, \overline X_0}{2} = 0.
\ee

\subsection{Type-I extrema}

Starting from a type-I extremum,
we use an $SU(2)$ rotation to place the whole VEV of $\Phi$ in $\phi_2$,
while $\phi_1$ has null VEV.
Then,
at a type-I extremum
\bs
\ba
\overline \varphi_\mathrm{I} &=& \left( \begin{array}{ccccc}
  0, & v_\phi, & 0, & \ldots, & 0 \end{array} \right)^T,
\\
\overline X_\mathrm{I} &=& \left( \begin{array}{cccc}
  v_\phi^2, & 0, & \ldots, & 0 \end{array} \right)^T.
\label{xI}
\ea
\label{typeIvac}
\es
Without loss of generality
we have assumed $v_\phi$ to be real.

At this extremum,
the stationarity conditions of Eq.~\eqref{stat_cond}
are trivially satisfied for all fields except for $\phi_2$.
For that field one has
\ba
\left.\frac{\partial V}{\partial \phi_2}
\right|_{\varphi = \overline \varphi_\mathrm{I}} = 0
&\Leftrightarrow&
\left( \mu_0^2 + \lambda_{00} v_\phi^2 \right) v_\phi = 0.	
\ea
Since we \emph{assume} $v_\phi$ to be nonzero,
we must have
\be
v_\phi^2 = - \frac{\mu_0^2}{\lambda_{00}}.
\label{vphi}
\ee
The conditions for $V$ to be bounded from below
require $\lambda_{00}$ to be positive.
Hence,
Eq.~\eqref{vphi} tells us that $\mu_0^2$ must be negative
for a type-I extremum to exist.
From Eq.~\eqref{mc09},
\be
\overline V_\mathrm{I} = \frac{L^T\, \overline X_I}{2}
= \frac{\mu_0^2 v_\phi^2}{2}
= \frac{- \left( \mu_0^2 \right)^2}{2 \lambda_{00}},
\ee
which is negative because $\lambda_{00} > 0$.
Therefore,
\emph{a type-I extremum has a lower value of the potential
than the type-0 extremum}:
\be
\label{0}
\overline V_0 > \overline V_\mathrm{I}.
\ee

We next search in the potential
for the terms quadratic in the fields of $\Psi_k$
when $\varphi = \overline \varphi_\mathrm{I}$.
We find
\begin{equation}
V \supset
\sum_{k=1}^N\, \sum_{I=-J_k}^{J_k}
\left( \mu_k^2 + \lambda_{0k} v_\phi^2 
- \frac{\lambda_{0k}^\prime}{2}\, I v_\phi^2
\right) \left| \psi_{k, I} \right|^2.
\end{equation}
Therefore, at this extremum,
the masses-squared of the complex scalar fields $\psi_{k, I}$ are given by
\be
m^2_{\psi_{k, I}} = \mu_k^2 + \left( \lambda_{0k}
- \frac{\lambda_{0k}^\prime}{2}\, I \right) v_\phi^2.
\label{mass2}
\ee
We assume that the type-I extremum is a local \emph{minimum} of the potential,
which means that all the $m^2_{\psi_{k, I}}$ are \emph{positive}.

\subsection{Type-III extrema}

Starting from the most general type-III extremum,
we use an $SU(2)$ rotation to place the whole VEV of $\Phi$ in $\phi_2$,
while $\phi_1$ has null VEV.
Then,
at a type-III extremum
\bs
\ba
\overline \varphi_\mathrm{III} &=& \left( \begin{array}{cccccccc} 
  0, 
  & v_b,
  & v_{\psi_{1, J_1}},
  & \ldots,
  & v_{\psi_{1, -J_1}},
  & v_{\psi_{2, J_2}},
  & \ldots,	
  & v_{\psi_{N, -J_N}}
\end{array} \right)^T,
\\
\overline X_\mathrm{III} &=& \left( \begin{array}{cccc}
  v_b^2, 
  & \sum_{I=-J_1}^{J_1} \left| v_{\psi_{1, I}} \right|^2, 
  & \ldots, 
  & \sum_{I=-J_N}^{J_N} \left| v_{\psi_{N, I}} \right|^2
\end{array} \right)^T.
\label{xIII}
\ea
\label{typeIIIvac}
\es
Without loss of generality, we have assumed $v_b$ to be real,
while the $v_{\psi_{k, I}}$ are, in general, complex numbers.

At a type-III extremum
\bs
\ba
\left.
\frac{\partial V}{\partial \phi_2}
\right|_{\varphi = \overline \varphi_\mathrm{III}} = 0
&\Leftrightarrow &
\left.
\frac{\partial V}{\partial F_0} \frac{\partial F_0}{\partial \phi_2}
\right|_{\varphi = \overline \varphi_\mathrm{III}}
+
\sum_{k=1}^N \left.
\frac{\partial V}{\partial F_{0k}}\, \frac{\partial F_{0k}}{\partial \phi_2}
\right|_{\varphi = \overline \varphi_\mathrm{III}} = 0
\\
&\Leftrightarrow &
v_b \left.
\frac{\partial V}{\partial F_0}
\right|_{\varphi = \overline \varphi_\mathrm{III}}  
+ \sum_{k=1}^N \lambda_{0k}^\prime \left.
\frac{\partial F_{0k}}{\partial \phi_2}
\right|_{\varphi = \overline \varphi_\mathrm{III}} = 0
\\
&\Leftrightarrow &
v_b \left(
\left.
\frac{\partial V}{\partial F_0}
\right|_{\varphi = \overline \varphi_\mathrm{III}} 
- \sum_{k=1}^N\, \sum_{I=-J_k}^{J_k}\,
\frac{\lambda_{0k}^\prime}{2}\, I \left| v_{\psi_{k, I}} \right|^2
\right) = 0,
\ea
\es
where we have used Eqs.~\eqref{dVdF0i} and~\eqref{13b}.
Since we \emph{assume} $v_b$ to be nonzero,
we must have
\be
\left.
\frac{\partial V}{\partial F_0}
\right|_{\varphi = \overline \varphi_\mathrm{III}} 
=
\sum_{k=1}^N\, \sum_{I=-J_k}^{J_k}\,
\frac{\lambda_{0k}^\prime }{2}\, I \left| v_{\psi_{k, I}} \right|^2.
\label{dVdF0}
\ee

\subsection{\label{Comparing extrema I and III}Comparing the type-I and type-III extrema}

Let us now suppose that the potential has two coexisting extrema
$\overline \varphi_\mathrm{I}$ and $\overline \varphi_\mathrm{III}$
of types I and III,
respectively.
Let us call the values of the scalar potential at those extrema $\overline V_\mathrm{I}$
and $\overline V_\mathrm{III}$,
respectively.
Using Eqs.~\eqref{dVdX} and~\eqref{mc09} one finds
\bs
\ba
\left. \frac{\partial V}{\partial X}
\right|_{\varphi = \overline \varphi_\mathrm{I}} \cdot
\left. X \right|_{\varphi = \overline \varphi_\mathrm{III}}
&=& \left( L^T + \overline X_\mathrm{I}^T M \right) \overline X_\mathrm{III}
\\
&=& 2\, \overline V_\mathrm{III}
+ \overline X_\mathrm{I}^T M \overline X_\mathrm{III},
\label{eq:XVp1}
\\
\left. \frac{\partial V}{\partial X}
\right|_{\varphi = \overline \varphi_\mathrm{III}} \cdot
\left. X \right|_{\varphi = \overline \varphi_\mathrm{I}}
&=& \left( L^T + \overline X_\mathrm{III}^T M \right) \overline X_\mathrm{I}
\\
&=& 2\, \overline V_\mathrm{I}
+ \overline X_\mathrm{III}^T M \overline X_\mathrm{I}.
\label{eq:XVp2}
\ea
\es
Since $M$ is symmetric,
$\overline X_\mathrm{I}^T M \overline X_\mathrm{III}
= \overline X_\mathrm{III}^T M \overline X_\mathrm{I}$.
Subtracting Eq.~\eqref{eq:XVp2} from Eq.~\eqref{eq:XVp1}
we find that the relative depth of the two extrema is
\cite{goncalves}
\be
\overline V_\mathrm{III} - \overline V_\mathrm{I} = \frac{1}{2} 
\left( 
\left. \frac{\partial V}{\partial X}
\right|_{\varphi = \overline \varphi_\mathrm{I}} \cdot
\overline X_\mathrm{III}
-
\left. \frac{\partial V}{\partial X}
\right|_{\varphi = \overline \varphi_\mathrm{III}} \cdot
\overline X_\mathrm{I}
\right).
\label{41}
\ee

Firstly,
from Eqs.~\eqref{dVdX},
\eqref{L},
\eqref{M},
and~\eqref{xI} one sees that
\be
\left. \frac{\partial V}{\partial X}
\right|_{\varphi = \overline \varphi_\mathrm{I}}
= L + \overline X_I^T M
=  \left( \begin{array}{c} 
\mu^2_0 + \lambda_{00} v_\phi^2 \\
\mu^2_1 + \lambda_{01} v_\phi^2 \\
\vdots \\
\mu^2_N + \lambda_{0N} v_\phi^2
\end{array} \right)
=  \left( \begin{array}{c} 
0 \\
\mu^2_1 + \lambda_{01} v_\phi^2 \\
\vdots \\
\mu^2_N + \lambda_{0N} v_\phi^2
\end{array} \right).
\ee
Utilizing Eq.~\eqref{xIII},
this means that
\be
\left. \frac{\partial V}{\partial X}
\right|_{\varphi = \overline \varphi_\mathrm{I}} \cdot
\overline X_\mathrm{III}
= \sum_{k=1}^N\, \sum_{I=-J_k}^{J_k}
\left( \mu_k^2 + \lambda_{0k}v_\phi^2 \right)
\left| v_{\psi_{k, I}} \right|^2.
\ee

Secondly,
it follows from Eqs.~\eqref{xI} and~\eqref{dVdF0} that
\be
\left. \frac{\partial V}{\partial X}
\right|_{\varphi = \overline \varphi_\mathrm{III}} \cdot
\overline X_\mathrm{I}
= 
v_\phi^2\, \left. \frac{\partial V}{\partial F_0}
\right|_{\varphi = \overline \varphi_\mathrm{III}} 
=
v_\phi^2\, \sum_{k=1}^N\, \sum_{I=-J_k}^{J_k}
\frac{\lambda_{0k}^\prime }{2}\, I \left| v_{\psi_{k, I}} \right|^2.
\ee

Thirdly,
by using Eqs.~\eqref{41} and~\eqref{mass2} one finds
\bs
\label{III}
\ba
\overline V_\mathrm{III} - \overline V_\mathrm{I} 
&=& 
\frac{1}{2}\, \sum_{k=1}^N\, \sum_{I=-J_k}^{J_k}
\left( \mu_k^2 
+ \lambda_{0k}v_\phi^2 
- \frac{\lambda_{0k}^\prime }{2}\, I\, v_\phi^2
\right)
\left| v_{\psi_{k,I}} \right|^2\\
&=&
\frac{1}{2}\, \sum_{k=1}^N\, \sum_{I=-J_k}^{J_k}
m^2_{\psi_{k,I}} \left| v_{\psi_{k,I}} \right|^2,
\ea
\es
which is positive because we have assumed that the masses-squared
$m^2_{\psi_{k,I}}$ are
positive, Q.E.D.

We conclude that \emph{a local minimum of type-I
has a lower value of the potential than any type-0 or type-III extremum},
\textit{cf.}\ Eqs.~\eqref{0} and~\eqref{III},
which means that,
either the global minimum of the potential is that type-I local minimum,
or it is a type-II local minimum
(if such a local minimum exists).
We refer to appendix \ref{sec:appA} for a pedagogical application 
of this theorem to a model with two $SU(2)$ doublets.

\section{\label{sec:conclusions}Summary} 

In this paper,
we have generalized a result demonstrated by FG~\cite{goncalves}.
They considered an $SU(2) \times U(1)$ gauge model
with an $SU(2)$ doublet $\Phi$,
an $SU(2)$ triplet $\Delta$,
and an additional symmetry
$\Phi \to \Phi \exp \left( i \varkappa \right),\
\Delta \to \Delta \exp \left( i \varpi \right)$;
we have allowed $\Delta$ to be \emph{any} $SU(2)$ multiplet,
not necessarily a triplet. 
As a matter of fact,
our theorem is valid for larger scalar sectors
with several $SU(2)$ multiplets,
provided the symmetries of the model
are such that the scalar potential (SP)
is of the form in Eq.~\eqref{SP}.
We wanted to check the conditions under which
a \emph{local} minimum of the SP
where only $\Phi$ has a nonzero VEV is the \emph{global} 
minimum of the potential.
We have proved that this local minimum of the SP
lies \emph{below} any extremum of the SP
where \emph{both} $\Phi$ and $\Delta$ have nonzero VEVs.
As FG rightly remarked,
this still leaves open the possibility that
the global minimum of the relevant SP is an extremum where
\emph{only} $\Delta$ has nonzero VEV.
Our result reduces the need for a full 
minimization of the SP of many Standard Model extensions,
as some vacuum configurations are immediately disfavored.
This is especially convenient in Dark Matter models
such as the $U(1)$-invariant 
2HDM
(see appendix \ref{sec:appA}).
This theorem may also prove useful in the Grand
Unified Theories based on the Lie groups $SU(5)$ and $SO(10)$ 
where one or more scalar quadruplets (and in some cases even quintuplets)
of $SU(2)$, with various hypercharges, may be present---depending
on the scalar multiplets used to effect the
spontaneous symmetry breaking.

\vspace*{5mm}

\begin{acknowledgments}
This work has been supported
by the Portuguese Foundation for Science and Technology (FCT)
through projects UIDB/00777/2020, UIDP/00777/2020, and 
through the PRR (Recovery and Resilience Plan),
within the scope of the investment ``RE-C06-i06 - Science
Plus Capacity Building", measure ``RE-C06-i06.m02 -
Reinforcement of financing for International Partnerships
in Science, Technology and Innovation of the PRR", under
the project with reference 2024.01362.CERN.
A.M. was additionally supported by FCT with PhD Grant No. 2024.01340.BD.
L.L.\ was furthermore supported by projects CERN/FIS-PAR/0002/2021
and CERN/FIS-PAR/0019/2021.
\end{acknowledgments}

\appendix

\section{\label{sec:appA}The $U(1)$-symmetric 2HDM}

Consider the simple case of two $SU(2)$ doublets: $\Phi$ and $\Psi$.
We write the field components of $\Phi$, $\Psi$, and their 
charge-conjugates $\widetilde{\Phi}$ and $\widetilde{\Psi}$,
respectively, as
\be
\Phi = \left( \begin{array}{c} a \\ b \end{array} \right),
\quad\quad
\Psi = \left( \begin{array}{c} c \\ d
\end{array} \right),
\quad\quad
\widetilde{\Phi} = \left( \begin{array}{c} b^* \\ -a^* \end{array} \right),
\quad\quad
\widetilde{\Psi} = \left( \begin{array}{c} d^* \\ - c^*
\end{array} \right).
\ee

We assume the scalar potential to be the one given in Eq.~\eqref{SP},
namely
\ba
\label{SPIDM}	
V &=&
\mu_0^2\, F_0 +
\mu_1^2\, F_1 +
\frac{\lambda_{00}}{2}\, F_0^2 +  
\frac{\lambda_{11}}{2}\, F_1^2 +  
\lambda_{01} F_0 F_1 +
\lambda_{01}^\prime F_{01} ,
\ea
where
\bs
\ba
F_0 &\equiv &
 \Phi^\dagger \Phi ,\\
F_1 &\equiv &
 \Psi^\dagger \Psi ,\\
F_{01} &\equiv &
\frac{1}{4}\left( |a|^2 - |b|^2 \right)
\left( |c|^2 - |d|^2 \right)+
\frac{1}{2}\left( ab^*c^*d + a^*bcd^*
\right).
\ea
\es

At a type-I extremum,\footnote{This 
vacuum configuration corresponds to the one used in the
$U(1)$-symmetric Inert Doublet Model~\cite{Ginzburg:2010wa,Jueid:2020rek}.}
we use an $SU(2)$ rotation to place the whole VEV of $\Phi$ in $b$,
leaving $a$ with null VEV. Explicitly, we write
\begin{equation}
\label{IDM_type1}
\langle \Phi \rangle = \left( \begin{array}{c} 0 \\ v_b \end{array} \right),
\quad \quad \quad 
\langle \Psi \rangle = \left( \begin{array}{c} 0 \\ 0
\end{array} \right),
\end{equation}
where $v_b$ is real.
Hence, at a type-I extremum,
the squared masses of the fields $c$ and $d$ read
\bs
\ba
m_c^2 &= \mu_1^2 + v_b^2
	\left( \lambda_{01} -\frac{1}{4} \lambda_{01}^\prime \right),\\
m_d^2 &= \mu_1^2 + v_b^2
	\left( \lambda_{01} +\frac{1}{4} \lambda_{01}^\prime \right).
\ea
\es
At a type-III extremum, the VEV of $\Phi$ can still be
written as in Eq.~\eqref{IDM_type1}, but now $\Psi$ may acquire
one of three vacuum configurations:
\begin{equation}
\langle \Psi \rangle_A = \left( \begin{array}{c} v_c \\ 0 \end{array} \right),
\quad \quad \quad 
\langle \Psi \rangle_B = \left( \begin{array}{c} 0 \\ v_d
\end{array} \right),
\quad \quad \quad 
\langle \Psi \rangle_C = \left( \begin{array}{c} v_c \\ v_d
\end{array} \right),
\end{equation}
with $v_c$ and $v_d$ being, in general, complex numbers.
According to Eq.~\eqref{III}, 
the relative depth of the scalar potential 
between the type-I extremum
and any of the three type-III extrema reads
\bs
\ba
\langle \Psi \rangle_A &:&
\overline V_\mathrm{III}^A - \overline V_\mathrm{I} 
= \frac{1}{2} m^2_{c} \left| v_{c} \right|^2,\\
\langle \Psi \rangle_B &:&
\overline V_\mathrm{III}^B - \overline V_\mathrm{I} 
= \frac{1}{2} m^2_{d} \left| v_{d} \right|^2,\\
\langle \Psi \rangle_C &:&
\overline V_\mathrm{III}^C - \overline V_\mathrm{I} 
= \frac{1}{2} m^2_{c} \left| v_{c} \right|^2 
+\frac{1}{2} m^2_{d} \left| v_{d} \right|^2.
\ea
\es

To compare these results with the literature, it is convenient
to write the scalar potential of Eq.~\eqref{SPIDM} 
in the language of the 2HDM, where the most general scalar potential
is usually written as \cite{Branco:2011iw}
\bs\ba
V &=& -\frac{1}{2} m_{11}^2 \Phi^\dagger \Phi
    -\frac{1}{2} m_{22}^2 \Psi^\dagger \Psi 
    - \left( m_{12}^2 \Phi^\dagger \Psi + \text{H.c.} \right) \\
&& + \frac{1}{2} \lambda_1 (\Phi^\dagger \Phi)^2 
    + \frac{1}{2} \lambda_2 (\Psi^\dagger \Psi)^2 
    + \lambda_3 (\Phi^\dagger \Phi)(\Psi^\dagger \Psi) 
    + \lambda_4 (\Phi^\dagger \Psi)(\Psi^\dagger \Phi) \\
&& + \left[ \frac{1}{2} \lambda_5 (\Phi^\dagger \Psi)^2 
    + \lambda_6 (\Phi^\dagger \Phi)(\Phi^\dagger \Psi) 
    + \lambda_7 (\Psi^\dagger \Psi)(\Phi^\dagger \Psi) + \text{H.c.} \right] .
\ea\es
Equation~\eqref{SPIDM} can be cast into this notation
by first setting $m_{12}^2 = 0$, 
$\lambda_{5,6,7} = 0$---this corresponds to the $U(1)$-symmetric 
2HDM~\cite{Branco:2011iw}---and then by 
performing the following substitutions
\bs
\ba
&&\mu_0^2 \rightarrow -\frac{1}{2} m_{11}^2,\quad
\mu_1^2 \rightarrow  -\frac{1}{2} m_{22}^2,\\
&&\lambda_{00}  \rightarrow  \lambda_1,\quad
\lambda_{11} \rightarrow  \lambda_{2},\quad
\lambda_{01} \rightarrow  \lambda_{3} + \frac{1}{2} \lambda_{4},\quad
\lambda_{01}^\prime \rightarrow  2 \lambda_{4}.
\label{subs}
\ea
\es

In Ref.~\cite{Ginzburg:2010wa}
the stability of the (inert) type-I vacuum against the type-III vacuum
$\left\langle \Psi \right\rangle_B$ has been studied.
In Eq.~(19), the authors state a condition 
that guarantees that 
$\overline V_\mathrm{III}^B - \overline V_\mathrm{I} > 0$, namely
\be
\overline V_\mathrm{III}^B - \overline V_\mathrm{I}
=
\frac{\left(m_{22}^2\lambda_1 - m_{11}^2(\lambda_3 + \lambda_4)\right)^2}
	{8 \lambda_1 \left( (\lambda_3 + \lambda_4)^2 - \lambda_1\lambda_2 \right)}.
\label{condition_IDM_ref}
\ee
However, realizing that
\bs
\ba
v_b
&=&
-\frac{\mu_0^2}{\lambda_{00}} = \frac{m_{11}^2}{2 \lambda_1},\\
v_d
&=&
\frac{\mu_0^2\left(\lambda_{01}+\frac{1}{4}\lambda_{01}^\prime\right)
 - \mu_1^2\lambda_{00}}
 {\lambda_{00}\lambda_{11}  -  \left(\lambda_{01}+\frac{1}{4}\lambda_{01}^\prime \right)^2} = 
\frac{m_{22}^2\lambda_1 - m_{11}^2(\lambda_3 + \lambda_4)}
	{2 \left(  \lambda_1\lambda_2  -  (\lambda_3 + \lambda_4)^2 \right)},
\ea
\es
the quantity in Eq.~\eqref{condition_IDM_ref} simply becomes
\be
\frac{\left(m_{22}^2\lambda_1 - m_{11}^2(\lambda_3 + \lambda_4)\right)^2}
	{8 \lambda_1 \left( (\lambda_3 + \lambda_4)^2 - \lambda_1\lambda_2 \right)}
=
\frac{1}{2} m_d^2 \left| v_d \right|^2,
\ee
which is necessarily positive because 
$m_d^2$ must be positive for the type-I extremum to be a local minimum
of the scalar potential.

This not only confirms the consistency of our results in the 
context of the $U(1)$-symmetric 2HDM
but also
highlights the usefulness of this theorem in reducing redundancy
when minimizing a scalar potential, 
as some vacuum configurations are immediately disfavored.

\end{document}